potmsb.tex0021263407453262232012034 0ustar  RiskaRiska
\documentclass[12pt]{article}
\begin{document}

\title{Dynamical Interpretation of the 
Nucleon-Nucleon Interaction and Exchange Currents
in the Large $N_C$ Limit}

\author{D.O. Riska\footnote{riska@pcu.helsinki.fi} 
\\ \vspace{0.3cm} \\
{\normalsize \it Helsinki Institute of Physics
and Department of Physics}\\
{\normalsize\it 00014 University of Helsinki, Finland} }

\date{}
\maketitle
\thispagestyle{empty}

\begin{abstract}

Expression of the nucleon-nucleon interaction to order
$1/N_C$ in terms of Fermi Invariants allows a dynamical
interpretation of the interaction and leads to a
consistent construction of the associated interaction
currents to order $1/N_C$. The numerically significant
components of 4 different modern realistic phenomenological
interaction models are shown to admit very similar meson
exchange interpretations in the large $N_C$ limit.
Moreover the ratio of the volume integrals of the
leading, next-to-leading and next-to-next leading
order terms in these interaction models is roughly
300:5-10:0.1, which corresponds fairly well to the
ratios of $1/N_C^2$ between the terms that would be
suggested by the $1/N_C$ expansion if $N_C=3$.
The $N_C$ dependence of the electromagnetic and
axial interaction currents that 
are associated with these interaction components
is derived and compared to that of the corresponding
single nucleon currents. 

\end{abstract}

\newpage

\section{Introduction}

The most promising $QCD-$based perspective on nuclear structure 
may very well be that provided by the large color number
($N_C$) limit, which allows a systematic series expansion
in $1/N_C$ for hadronic observables, where the first few
terms appear to capture the key phenomenological features
of the structure of the baryons \cite{Jenkins}. The leading terms of 
the $1/N_C$ series expansion of the components of the 
nucleon-nucleon have been shown to correspond well with
the strongly coupled boson exchange terms in a phenomenological
boson exchange model for the interaction \cite{Kaplan}. While
the large $N_C$ limit itself cannot determine the radial
behavior of the potential components, this result
nevertheless provides a systematic, if approximate connection 
between nuclear phenomenology and $QCD$.

Here it will be shown that in fact all the 
most commonly employed modern 
realistic 
phenomenological interaction models a) allow an interpretation in terms 
of phenomenological boson exchange and b) that the numerically
significant (or well determined) components of these 
interactions correspond very well to the leading terms
in the $1/N_C$ expansion of the nucleon-nucleon interaction
components. The next-to-leading order terms are are shown
to be numerically very small in comparison to the terms
of leading order. The ratio of the volume integrals 
of the leading, next-to-leading and next-to-next leading
order terms in these interaction models is roughly
300:5-10:0.1, which corresponds fairly well to the
ratios of $1/N_C^2$ between the terms that would be
suggested by the $1/N_C$ expansion if $N_C=3$.

Finally the corresponding $1/N_C$ scaling factors  
of the associated two-nucleon exchange (or
interaction) currents are derived. The axial exchange
charge operator that is associated with long
range pion exchange is shown to scale as $N_C^0$, as
the single nucleon axial charge, whereas those
that are associated with the Fermi invariant
components of the interaction scale as
$1/N_C$ or higher powers in $1/N_C$. The
electromagnetic exchange currents scale with
two additional powers of $1/N_C$ in comparison
with the single nucleon current. The axial
exchange current scales as $1/N_C^3$,
and is thus smaller by $1/N_C^4$ than the
axial current of the nucleon. These scaling
factors correspond well with established
nuclear phenomenology.

The key to the dynamical interpretation of a phenomenological
interaction model is to express it in terms of the 5 Fermi
invariants $S,V,T,A,P$. These invariants form a unique set,
that is independent of particle momenta, and allow an
interpretation of the interaction in terms of linear combinations 
of scalar, vector, axial vector and pseudoscalar exchange 
mechanisms, without the need for explicit specification of 
the radial behavior of the
potential functions. 
Moreover all of these invariants
define unique and consistent interaction current operators 
\cite{Mariana,Tsushima}, for which 
$1/N_C$ expansions follow from those
of the corresponding potentials, as will be shown below.

The utility of the large $N_C$ limit of $QCD$ in nuclear
phenomenology has emerged only gradually, the first
indication being the a priori surprising phenomenological 
quality of the description of selected observables of
nucleons, hyperons and nuclei in terms of Skyrme's
topological soliton model \cite{Skyrme} and its
generalizations. These models represent
one realization of the large $N_C$ limit, in which
the baryons appear as topologically stable solutions
of nonlinear chiral meson field theories \cite{Nyman}.   
In particular it was found that Skyrme's product ansatz
for the baryon number 2 system led to a phenomenologically
satisfactory description of the longest range components
of the isospin dependent part of the nucleon-nucleon
interaction \cite{Skyrme2,Nyman2}. This result is in
fact very natural as  these components of the interaction
are of leading order in the $1/N_C$ expansion 
for which the Skyrme model 
should give generically adequate results.

The modern phenomenological nucleon-nucleon interaction
models that are considered here are the 
$V18$ \cite{V18},
the $CD-Bonn$ \cite{Mach}, the $Nijmegen(93)$ 
\cite{Swart} and the $Paris$ \cite{Paris}
potentials. All of these contain a long range pion
exchange tail, and have mostly or purely phenomenological
short and intermediate range terms. Their components in the
$S,V,T,A,P$ representation are nevertheless remarkably
similar in strength, and if parametrized in terms
of single meson exchange give rise to ``effective''
meson-nucleon coupling strengths, which also are
very similar.

In the large $N_C$ limit the transformation between the 
phenomenological potential components given in the spin 
representation and the representation in terms of Fermi
invariants reveals a new result in that
leading order parts of the isospin dependent scalar and 
vector potentials have to be equal in magnitude 
with opposite sign. If interpreted in terms
of meson exchange this suggests that $a_0$ and
$\rho$ meson exchange have equal strength
in the large $N_C$ limit.

In section 2 the expressions for the Fermi
invariant decomposition for the large $N_C$
limit of the nucleon-nucleon interaction
are derived.
In section 3 the corresponding components
for the
4 interaction models considered are calculated.
All of these are then shown to admit a meson
exchange interaction with very similar coupling
strengths. Their well determined components are
then shown to be consistent with the interaction
form suggested by the leading terms in the
$1/N_C$ expansion. In sections 4 and 5 the 
$N_C$ scaling factors of the corresponding 
electromagnetic and axial
interaction current operators are derived. A 
concluding discussion completes the paper.

\section{The Nucleon-Nucleon Interaction in terms
of Fermi Invariants in the large $N_C$ Limit}
\label{2sec}

\subsection{Phenomenological interactions}

The nucleon-nucleon interaction is commonly expressed
in terms of the following set of Galilean invariant
spin- and isospin operators:
\begin{equation} 
V_{NN}=\sum_{i}^5 [\tilde v_j^+ + \tilde v_j^- \vec\tau^1\cdot
\vec\tau^2]\,\tilde\Omega_j,
\label{e1}
\end{equation} 
where the coefficients $\tilde v_j^\pm$ are scalar
functions and
the spin operators $\tilde\Omega_j$ are defined as
\begin{eqnarray}
&&\tilde\Omega_C=1,\quad\tilde\Omega_{LS}=\vec L
\cdot\vec S,\quad\tilde\Omega_T=S_{12},\nonumber\\
&&\tilde\Omega_{SS}=\vec\sigma^1\cdot\vec\sigma^2,\quad
\tilde\Omega_{LS2}={1\over 2}\{\vec\sigma^1\cdot\vec L,
\vec\sigma^2\cdot\vec L\}_+\quad.
\label{e2}
\end{eqnarray} 
Because the little group of transformations, which
leave the square of the 4-velocity of the two-particle
system invariant is the same for Galilean and
Poincar\'e transformations, the interaction
(\ref{e1}) is invariant under Poincar\'e
transformations.

The power of the leading term in the
$1/N_C$ expansion for these
potential components $\tilde v_j^\pm$, and therefore their
order in $1/N_C$, has been shown to be
the following \cite{Kaplan}:
\begin{eqnarray}
&&\tilde v_C^+\tilde\Omega_C,\quad 
\tilde v_T^-\tilde\Omega_T,\quad 
\tilde v_{SS}^-\tilde\Omega_{SS}
\quad\sim\quad {\cal O}(N_C),
\nonumber\\
&&\tilde v_C^-\tilde\Omega_C,\quad 
\tilde v_2^\pm\tilde\Omega_{LS},\quad 
\tilde v_T^+\tilde\Omega_T,
\quad \tilde v_{SS}^+ \tilde\Omega_{SS},
\quad \tilde v_{LS2}^- \tilde\Omega_{LS2}
\quad\sim\quad {\cal O}(1/N_C),
\nonumber\\
&&\tilde v_{LS2}^+ \tilde\Omega_{LS2}
\quad \sim\quad {\cal O}(1/N_C^3).
\label{e3}
\end{eqnarray} 
A simple derivation of these scaling relations is
given below.
The numerically significant, or ``large'', potential
components should therefore, on the basis of their
order in $1/N_C$, be 
the isospin independent
central, and isospin dependent spin-spin and
tensor components of the potential. This corresponds
completely to established nuclear structure 
phenomenology, as the first of these terms corresponds
to the short range repulsive and intermediate range
attractive force components, and the latter two
correspond to components of the long range pion 
exchange interaction. These are also the force
components that are readily derived in the
Skyrme model \cite{Skyrme2}.

While the 5 linearly independent potential invariants
$\tilde\Omega_j $ are sufficient to describe the two-
nucleon amplitude for nucleons on their mass shell,
they are not dynamically transparent. For a
dynamical interpretation, which allows a connection
to field theory, reexpression of the interaction in
terms of the 5 Fermi invariants $S,V,T,A,P$ is
required. These are defined as \cite{ggmw}
\begin{eqnarray}
&&S=1,\quad V=\gamma_\mu^1 \gamma_\mu^2,\quad
T={1\over 2} \sigma_{\mu\nu}^1\sigma_{\mu\nu}^2,
\nonumber\\
&&    
A=i\gamma_5^1\gamma_\mu^1 \,\, i\gamma_5^2\gamma_\mu^2,
\quad P=\gamma_5^1\gamma_5^2.
\label{e4}
\end{eqnarray}
The nucleon-nucleon interaction, when expressed
in terms of the Fermi invariants, takes the form
\begin{equation}
V_{NN}=\bar u(p_1')\bar u(p_2')
[v_j^+ + v_j^-\,\vec\tau^1\cdot\vec\tau^2]\,F_j\,
u(p_1)u(p_2),
\label{e5}
\end{equation}
where the $F_j$, $j=1...5$ represent the Fermi 
invariants in the order $S,V,T,A,P$.

To find the relation between the potential functions
$v_j$ and the nonrelativistic potential components
$\tilde v_j$ in (\ref{e1}) the spin representation the Fermi 
invariants is required: 
\begin{eqnarray}
&&\Omega_C=1,\quad \Omega_{LS}={i\over 2}(\vec\sigma^1
+\vec\sigma^2)\cdot\vec n,
\nonumber\\
&&
\Omega_T=\vec\sigma^1\cdot\vec\sigma^2\,k^2
-3\vec\sigma^1\cdot\vec k\,\vec\sigma^2\cdot \vec k,
\quad \Omega_{SS}=\vec\sigma^1\cdot\vec\sigma^2,
\nonumber\\
&&
\Omega_{LS2}=\vec\sigma^1\cdot\vec n \,\vec\sigma^2\cdot\vec n.  
\label{e6}
\end{eqnarray}  
Here the momenta $\vec k,\vec P$ and $\vec n$ are
defined as $\vec k=\vec p\,'-\vec p$, $\,$
$\vec P=(1/2)(\vec p\,'+\vec p)$ and
$\vec n=\vec k\times\vec P$, where $\vec p\,'$ and
$\vec p$ are the final and initial relative momenta.
The linear relation between the spin operators $\Omega_j$
and the Fermi invariants $F_j$ is given (to order
$1/m_N^2$) in ref. \cite{Tsushima}. Since the nucleon
mass $m_N\sim {\cal O}(N_C)$, only the leading terms in
$1/m_N^2$ are needed for the present large $N_C$
limit considerations. 

The momentum space representation of the interaction
(\ref{e1}) is
\begin{equation} 
V_{NN}=\sum_{i}^5 [w_j^+ + w_j^- \vec\tau^1\cdot
\vec\tau^2]\,\Omega_j.
\label{e7}
\end{equation} 
Note that $ w_{LS2} \Omega_{LS2}$ only represents part of the
complete Fourier transform of $\tilde v_{LS2}\tilde\Omega_{LS2}$, 
which also contains a linear combination of a
term with the operators
$\vec\sigma_1\cdot\vec P\,\vec\sigma_2\cdot\vec P -\vec P\,^2
\vec\sigma_1\cdot\vec\sigma_2\,$, $\Omega_3$ and $\Omega_4$
\cite{Swart} .
These additional terms are of the same order in
$1/N_C$ as the term $w_{LS2}^\pm\Omega_{LS2}$.

For a local potential one has explicitly \cite{Blunden}
\begin{eqnarray}
&& w_C^\pm (k) =4\pi\int_0^\infty dr r^2 j_0(kr) 
\tilde v_C^\pm (r),
\nonumber\\
&&w_{LS}^\pm (k)=-{4\pi\over k}\int_0^\infty  dr r^3 j_1(kr) 
\tilde v_{LS}^\pm (r),
\nonumber\\
&&w_T^\pm (k)={4\pi\over k^2}\int_0^\infty  dr r^2 j_2(kr) 
\tilde v_T^\pm (r),
\nonumber\\
&&w_{SS}^\pm (k)=4\pi\int_0^\infty  dr r^2 j_0(kr) 
\tilde v_{SS}^\pm (r), \nonumber\\
&&w_{LS2}^\pm (k)=-{4\pi\over k^2}\int_0^\infty  dr r^4 j_2(kr) 
\tilde v_{LS2}^\pm (r).
\label{e8}
\end{eqnarray}

\subsection{Order in $N_C$ of the potential components}

The $N_C$ dependence of the components of the potential
(\ref{e7}) may be inferred directly by quark model
considerations. Consider the single quark operators
$1,\sigma_j,\,\tau_k,\,\sigma_j\tau_k$. Matrix elements of 
the sum over $N_C$ such quark operators in nucleon
states then depend on $N_C$ as follows \cite{Wirzba}:
\begin{eqnarray} 
&&\langle N\vert \sum_{q=1}^{N_C} 1^q \vert
N \rangle \sim N_C,
\quad
\langle N\vert \sum_{q=1}^{N_C} \sigma_j^q\vert
N \rangle \sim N_C^0
\nonumber\\
&&\langle N\vert \sum_{q=1}^{N_C} \tau_k^q \vert
N \rangle \sim N_C^0
\quad
\langle N\vert \sum_{q=1}^{N_C} \sigma_j^q\tau_k^q\vert
N \rangle \sim N_C.
\label{e9}
\end{eqnarray}
In the large $N_C$ limit the baryon-baryon interaction
may be interpreted as meson exchange, since the
gluon lines may be replaced by $q\bar q$ lines in all
surviving (planar gluon) diagrams. The meson-baryon couplings are
proportional to the quark operator matrix elements
above and inversely proportional to the meson
decay constants $f_M$, which scale like $\sqrt{N_C}$
\cite{Jenkins}. 
Application of the scaling relations (\ref{e9})
and multiplication by $1/f_M^2$ to the
two nucleon system directly implies that
\begin{eqnarray}
&&w_C^+\Omega_C,\quad w_T^-\Omega_T,\quad w_{SS}^-\Omega_{SS}
\quad\sim \quad{\cal O}(N_C),
\nonumber\\
&& w_C^-\Omega_C,\quad w_T^+\Omega_T,\quad w_{SS}^+\Omega_{SS}
\quad\sim\quad {\cal O}(1/N_C).
\label{e10}
\end{eqnarray}
The order in $N_C$ of the spin-orbit interaction
may be found as follows. The non-local momentum
operator $\vec P$ in the spin-orbit interaction
operator $\Omega_{LS}$ always appears in the combination 
$\vec P/m_N$, where $m_N\sim N_C$. The isospin
independent term $w_{LS}^+\Omega_{LS}$ contains the
spin-operator of one nucleon in combination with
$\vec P/m_N$, and therefore the vertex scales as
$1/N_C$. If the quark coupling at the other nucleon
line is $\sim 1^q$, it follows that 
$w_{LS}^+\Omega_{LS}\sim {\cal O}(1/N_C)$
once the overall factor $1/f_M^2$ is taken into
account. For the isospin dependent spin-orbit
interaction the scaling factor or the vertex
with a spin-operator multiplying $\vec P/m_N$ and
an isospin operator is $N_C^0$. In this case the
vertex at the other nucleon line also contains an
isospin factor, so it also scales as $N_C^0$. Thus,
it follows after multiplication with $1/f_M^2$ that
$w_{LS}^-\Omega_{LS}\sim {\cal O}(1/N_C)$. 

To find the $N_C$ scaling of the isospin independent quadratic 
spin-orbit interaction one notes that there is a combination
of a spin operator with $\vec P/m_N$ at both nucleon
lines, and thus the scaling factor is $1/N_C^2$, which
after multiplication with $1/f_M^2$ gives rise to
the scaling $w_{LS2}^+\Omega_{LS2}
\sim {\cal O}(1/N_C^3)$. 
The corresponding
scaling factor for the isospin independent 
quadratic spin-orbit interaction is
$w_{LS}^-\Omega_{LS2}\sim {\cal O}(1/N_C)$, because the 
isospin operators at
both nucleon lines bring an additional factor $N_C$ each.
This completes the derivation of the Manohar-Kaplan
scaling relations (\ref{e3}) \cite{Kaplan}.

The linear relation between the Fermi invariant potential 
components $v_j^\pm$ and the components $w_j^\pm$
is given in ref. \cite{Tsushima}. In the present
application, only the leading terms in $1/m_N^2$
(i.e. in $1/N_C^2$) have to be retained in the
transformation matrix. 

To leading order in $1/N_C$ the Fermi invariant potential 
components $v_j^\pm$ are then given as:
\begin{eqnarray}
&& v_S^+ ={3\over 4} w_C^+ -{m_N^2\over 2}w_{LS}^+,
\nonumber\\
&& v_S^- = -{m_N^2\over 2}w_{LS}^-
+{k^2\over 4}w_T^- +{1\over 4} w_{SS}^-
+{m_N^2 k^2\over 4} w_{LS2}^-,
\nonumber\\
&&v_V^+ = {1\over 4} w_C^+ +{m_N^2\over 2}w_{LS}^+,
\nonumber\\
&&v_V^- = {m_N^2\over 2}w_{LS}^- -{k^2\over 4}w_T^-
-{1\over 4} w_{SS}^- -{m_N^2 k^2\over 4} w_{LS2}^-,
\nonumber\\
&&v_T^+ = {k^2\over 32m_N^2}w_C^+ + {k^2\over 16}w_{LS}^+
+{k^2\over 2}w_T^+ +{1\over 2}w_{SS}^+
+{m_N^2 k^2\over 4}w_{LS2}^+,
\nonumber\\
&&v_T^- = {k^2\over 2}w_T^- +{1\over 2}w_{SS}^-
+{m_N^2 k^2\over 4}w_{LS2}^-,
\nonumber\\
&&v_A^+ = {k^2\over 32 m_N^2}w_C^+ + {k^2\over 16}w_{LS}^+
+{k^2\over 2}w_T^+ +{1\over 2}w_{SS}^+ 
-{m_N^2 k^2\over 4}w_{LS2}^+,
\nonumber \\
&&v_A^- = {k^2\over 2} w_T^- + {1\over2}w_{SS}^-
-{m_N^2 k^2\over 4}w_{LS2}^-,
\nonumber\\
&&v_P^+ = {1\over 4}w_C^+ +{m_N^2\over 2}w_{LS}^+
+12 m_N^2 w_T^+, 
\nonumber\\
&&v_P^- = 12 m_N^2 w_T^-\, .
\label{e11}
\end{eqnarray}
From these expressions it follows that both the isospin
independent and the isospin dependent scalar 
and vector potentials
$v_1^\pm\, S,\,v_2^\pm\, V$ are $\sim{\cal O}(N_C)$.
The isospin dependent tensor and
axial vector potentials $v_T^-\, T,\,v_A^-\, A$
are $\sim{\cal O}(N_C)$, whereas the corresponding
isospin independent interaction components
$v_T^+\, T,\,v_A^+\, A$ are $\sim{\cal O}(1/N_C)$.
The orders of the pseudoscalar
potentials are $v_P^+\, P\sim N_C$ and
$v_P^-\, P\sim N_C^3$. All the coefficients in the 
expression of the pseudoscalar 
operator $P$ in terms of the spin operators $\Omega_j$ 
are $\sim 1/m_N^2$ \cite{Tsushima}. The 
corresponding potentials in the spin
representation are therefore  
smaller by $1/N_C^2$ and as a consequence the highest
order in $N_C$ of the potentials components in the
spin representation is ${\cal O}(N_C)$.
This becomes explicit if the pseudoscalar 
invariant
$P$ is replaced by the on-shell equivalent pseudovector
invariant $P'$:
\begin{equation}
P'={1\over 4 m_N^2}\gamma_\mu^1 k_\mu\,\gamma_5^1\,\,
\gamma_\nu^2 k_\nu\,\gamma_5^2,
\label{e12}
\end{equation}
which has the form required by spontaneous breaking of
the approximate chiral symmetry of $QCD$. Because the
pseudovector invariant builds in the required pair
suppression in the pion-nucleon amplitude, and vanishes
with the 4-momentum of the exchanged system, it is
simpler to apply in relativistic approaches to
nuclear reaction phenomenology \cite{wallace}.
These results are summarized in
Table 1.

\begin{table}
\begin{center}
\begin{tabular}{|c|c|c|c|c|c|}
 \hline

&&&&&\\ 
Isospin & $S$ & $V$ & $T$ & $A$ & $P$ \\
&&&&&\\
 \hline     
&&&&&\\
$ 1$ & $\,\,\, N_C\,\,\,$ 
& $\,\,\, N_C\,\,\, $ & $1/N_C $ 
& $1/N_C$ & $N_C$
\\ 
&&&&&\\
\hline
&&&&&\\
$\vec \tau^1\cdot \vec \tau^2$& $\,\,\, N_C\,\,\,$
& $\,\,\,N_C\,\,\,$
& $\,\,\,N_C\,\,\,$
& $\,\,\,N_C\,\,\,$& $\,\,\,N_C^3\,\,\,$ 
\\
&&&&&\\
 \hline
\end{tabular}
\caption{Order of the leading term in the
$1/N_C$ expansion of the Fermi invariant potential
components.}
\end{center}
\end{table}

\subsection{Dynamical interpretation of the potential components}

The components of highest order in $1/N_C$ of the Fermi 
invariant potential components in Table 1 admit a direct 
dynamical interpretation in terms of meson exchange 
between the nucleons.

The only isospin independent potential components
that are of order $N_C$ (i.e. ``large'') are the
scalar and vector components $v_S^+$ and $v_V^+$
as well as the pseudoscalar component $v_P^+$.
The first one of these corresponds to the largest
component in the two-pion exchange interaction
between nucleons \cite{lomon,chemtob}, which commonly 
is modeled
in terms of a strong scalar (``$\sigma$'') meson
exchange mechanism \cite{brown,erkelenz}. 
This potential component gives rise to an attractive
interaction at intermediate range and the main agent
for nuclear binding. The second corresponds to
the short range repulsion between nucleons,
and is commonly modeled in terms of an
$\omega$ meson exchange interaction, with
an overstrength effective $\omega$ nucleon
coupling constant \cite{erkelenz}. The last
(pseudoscalar) term, which may be interpreted
as $\eta$ meson exchange is large in the earlier
potential models, but very small in the most
recently developed models. 

Among the isospin dependent Fermi invariant 
potential components in Table 1 the pseudoscalar
component is of
order $N_C^3$ and are thus the largest of
all terms in the $1/N_C$ counting scheme. This
strong pseudoscalar exchange component
$v_P^-$ is immediately interpretable in terms
of the long range pion exchange interaction
between nucleons, which is expectedly ``strong'', 
more because of its long range than because of the
strength of the pion-nucleon coupling.

That the other isospin dependent Fermi invariant
potential components are of order $N_C$
also corresponds to established nucleon-nucleon
phenomenology. To see this, it is worth noting
that an isospin 1 vector meson ($\rho$) exchange
interaction may be expressed in terms of Fermi
invariants as follows \cite{Blunden}:
\begin{eqnarray} 
&&(\gamma_\mu^1-{\kappa\over 2 m_N}\sigma_{\mu\nu}^1\,k_\nu)
(\gamma_\mu^2+{\kappa\over 2 m_N}\sigma_{\mu\alpha}^2
\,k_\alpha)
={\kappa\over m_N^2}P_\mu^1P_\mu^2 S +(1+\kappa)V
\nonumber\\
&& -{\kappa (1+\kappa) k^2\over 4 m_N^2} T
+{\kappa(1+\kappa)\over m_N^2}P_\mu^1 P_\mu^2 P.
\label{e13}
\end{eqnarray} 
Here to leading order in $1/m_N^2$ (or $1/N_C^2$):
\begin{equation}
P_\mu^1 P_\mu^2=-(m_N^2+{k^2\over 4}+2 \vec P^2).
\label{e14}
\end{equation}
Phenomenological boson exchange interaction models
typically contain a $\rho$ meson
exchange interaction, where the ``effective''
tensor coupling of the $\rho$ meson to nuclei
is large $\kappa\sim 7$ \cite{erkelenz} even though
the vector coupling is small. Because of the
large tensor coupling such an isospin dependent
vector meson exchange interaction contributes
strongly to all the Fermi invariant potential
components, except the axial vector invariant. 
The order $N_C$ of the potential
components $v_S^-,\,v_V^-,\, v_T^-$ thus
may be interpreted in terms of a strong $\rho$ meson
exchange (or $\rho$ exchange-like) interaction.
The isospin independent vector meson ($\omega$)
exchange in contrast only contributes strongly
to the vector $V$ invariant potential, because
of the small $\omega-$nucleon tensor
coupling.  

While the isospin dependent pseudoscalar potential
$v_P^- \, P$ is of order $N_C^3$ the corresponding
isospin independent pseudoscalar exchange potential
$v_P^+ \, P$ is only of order $N_C$. This corresponds
to the phenomenological finding in boson exchange
models for the nucleon-nucleon interaction that the
$\eta$ meson exchange term is mostly much weaker than
the $\pi$ meson exchange interaction component.

\section{Large $N_C$ components of phenomenological
interaction models}
\label{3sec}

\subsection{The scalar potential}
\label{3seca}

For the visualization of the Fermi invariant
potential components it is useful to plot the
potential components in configuration space.
The expression for the isospin independent 
and isospin dependent scalar
potentials in configuration space are obtained from
(\ref{e11}) as
\begin{eqnarray}
&&v_S^+(r) ={3\over 4}\,\tilde v_C^+(r) +{m_N^2\over 2}
\int_r^\infty dr' r'\,\tilde v_{LS}^+ (r'),
\nonumber\\
&& v_S^-(r) = {m_N^2\over 2}\int_r^\infty  
dr' r'\,\tilde v_{LS}^-(r')
-{1\over 4}\{\tilde v_T^-(r)-3\int_r^\infty dr'\,{\tilde v_T^-(r')
\over r'}\} 
+{1\over 4}\,\tilde v_{SS}^-(r)
\nonumber\\
&&-{m_N^2 \over 4}\{r^2 \tilde v_{LS2}^- (r) -3
\int_r^\infty dr' r'\, \tilde v_{LS2}^-(r')\} ,
\label{e15}
\end{eqnarray}

Here, and below, only terms of leading order in $1/N_C$ have
been retained, so that all terms on the r.h.s. are
of the same order in $1/N_C$.
These potential components as obtained from the phenomenological
$V18$ \cite{V18},
the $CD-Bonn$ \cite{Mach}, the $Nijmegen(93)$ 
\cite{Swart} and the $Paris$ \cite{Paris} potentials
are shown in Figs. 1 and 2.

In the case of the $V18$ potential, the operator form
of which contains a quadratic spin-orbit interaction
of the form $(\vec L\cdot \vec S)^ 2$, we employ the
relation 
\begin{equation}
(\vec L\cdot \vec S)^2=2\vec L^2 + 2\tilde\Omega_{LS2}
-2\tilde\Omega_{LS}
\label{e16}
\end{equation}
to reduce the interaction to standard form (\ref{e1})
in combination with terms of the form $\vec L^2$.
As the order of the remaining terms 
$\vec L^2$, $(\vec\tau^1
\cdot\vec\tau^2)\vec L^2$, $(\vec\sigma^1
\cdot\vec\sigma^2)\vec L^2$ and
$(\vec\tau^1\cdot\vec\tau^2) 
(\vec\sigma^1\cdot\vec\sigma^2)\vec L^2$ in the
$V18$ potential may, by the
argument in section (2.2), be shown to be at most of order
$1/N_C^{-1}$, $1/N_C^{-3}$, $1/N_C^{-3}$ and
$1/N_C^{-1}$, respectively (once multiplied by
$1/f_M^2$), we shall not have to retain them in the
leading order analysis here. The spin-orbit 
($\tilde \Omega_{LS}$) and quadratic spin-orbit
($\tilde\Omega_{LS2}$) interaction
terms in the $V18$ interaction 
shall however be included with
the corresponding terms in the expressions (\ref{e15}).

The $CD-Bonn$ interaction model is a single boson
exchange model that is expressed 
in terms of Fermi invariants and in the case of
vector meson exchange in the form, which 
is reducible to Fermi invariants
by eqn. (\ref{e13}). 
Both the $Nijmegen(93)$ and the (parametrized) $Paris$ 
interaction models are expressed in the
standard form (\ref{e1}). The central components of
the $Paris$ potential do in addition contain
velocity dependent terms with the operator
form $\vec P^2$. As these, by the argument in section
(2.2) are down by two orders in $1/N_C$ in comparison
to the local terms, they are not considered here.

In Table 2 are listed the volume integrals of the
the leading terms in the 
$1/N_C$ expansion of the Fermi invariant potential
components of the phenomenological interaction
models. In the case of the isospin independent
scalar interaction these volume integrals are all 
of the order 10 fm$^2$, ranging from -6.3 fm$^2$
for the $Paris$ potential to -13.5 for the
$CD-Bonn$ potential. 

If the isospin independent scalar potential 
component is interpreted as due to a single
scalar meson exchange interaction, the volume
integral would equal $-g_S^2/m_S^2$, where
$g_S$ is the scalar meson -- nucleon
coupling constant, and $m_S$ the scalar meson
mass. The mass of the lowest scalar meson
($\sigma$ or $f_0(400-1200)$) is not well
established \cite{PDG}, but it is typically
taken to be of the order 550 - 750 MeV in phenomenological
nucleon-nucleon interaction models \cite{erkelenz} .
If $m_S$ is taken to be 600 MeV, within this
range, the value of the effective scalar meson coupling
constant for these interactions range between 7.6
($Paris$) and 11.2 ($CD-Bonn$) (Table 3).
These ``effective'' scalar meson coupling
constant values should be viewed as lower limits,
as any short range form factors multiplying the
scalar meson exchange interaction would increase
these values. It is striking how closely similar
values obtain for the ``effective'' scalar meson
coupling constants with the 4 different interaction
models, two of which do not contain any scalar
meson like term in their parametrized forms.

The isospin independent scalar interaction components
of the 4 interaction models are rather similar
for nucleon separations beyond 0.5 fm. The radial
shapes differ considerably at short distances,
however, ranging from attractive to repulsive.
The two boson exchange models ($Nijmegen(93)$ and
$CD-Bonn$ are repulsive by construction. The
more phenomenological $V18$ and $Paris$ interaction
models are repulsive at short distance, a feature
which cannot be interpreted in terms of meson
exchange, and which otherwise only is known
from the Skyrme model \cite{Nyman}.

The isospin dependent scalar interaction components
of the 4 phenomenological interaction models are 
shown in Fig.2. These interaction components
are of order $1/N_C$ and should therefore be weaker
by and order of magnitude than the isospin 
independent scalar interaction components. This
is in fact also revealed by comparison of Figs.
1 and 2. The volume integrals of these
interaction components range between --3.2 
and --4.4 fm$^2$, as shown in Table 2, and are
thus about a factor 3-4 smaller than those
of the corresponding isospin independent
interactions. 

In a meson exchange interpretation the 
isospin dependent scalar interaction arises
from exchange of scalar mesons with isospin 1.
The lightest of these is the $a_0(980)$
In a simple meson exchange model the volume
integrals of these interactions should
equal $-g_{a_0}^2/m_{a_0}^2$, where 
$g_{a_0}$ is the $a_0(980)-$nucleon coupling
constant and $m_{a_0}$ is the mass of the
$a_0(980)$. If the value of the ``effective''
$a_0(980)$ nucleon coupling constant is
extracted from these volume integrals, the
values range from $g_{a_0} = 9.0$ to  
$g_{a_0} = 10.4$ (Table 3). In this case
all the interaction models give very
similar values for this coupling constant.

\begin{table}
\begin{center}
\begin{tabular}{|c|c|c|c|c|c|}
 \hline

&&&&&\\ 
Component & $V18$ & $CD-Bonn$ & $Nijmegen(93)$ & 
$Paris$& ${\cal O}(N_C)$ \\   
&&&&&\\
 \hline     
&&&&&\\
 $v_S^+$ &  -8.7 
& -13.5 & -10.3
& -6.3& $N_C$\\ 
&&&&&\\
\hline
&&&&&\\
$v_S^-$& -3.2
& -3.3
& -3.3
& -4.4 &$N_C$
\\
&&&&&\\
 \hline\hline
&&&&&\\

$v_V^+$& 9.4
& 11.6
& 8.7
& 10.2 &$N_C$
\\
&&&&&\\
 \hline
&&&&&\\
$v_V^-$& 3.2
& 3.2
& 2.9
& 4.4 &$N_C$
\\
&&&&&\\
 \hline\hline
&&&&&\\

$v_T^+$& -0.1
& 0.0
& 0.0001
& 0.03&$1/N_C$
\\
&&&&&\\
 \hline
&&&&&\\
$v_T^-$& 0.6
& 0.1
& 0.001
& 0.46 &$N_C$
\\
&&&&&\\
 \hline \hline
&&&&&\\

$v_A^+$& -0.1
& 0
& 0
& 0.03&$1/N_C$
\\
&&&&&\\
 \hline
&&&&&\\
$v_A^-$& 0.6
& 0
& 0
& 0.46 &$N_C$
\\
&&&&&\\
 \hline \hline
&&&&&\\

$v_P^+$& 9.8
& 0.0
& 0.35
& 18.0 &$N_C$
\\
&&&&&\\
 \hline
&&&&&\\
$v_P^-$& 360
& 338
& 323
& 352 &$N_C^3$
\\
&&&&&\\
 \hline
\end{tabular}
\caption{Volume integrals (in fm$^2$) of
the leading terms in the 
$1/N_C$ expansion of the Fermi invariant potential
components phenomenological interaction
models.}
\end{center}
\end{table}

\begin{table}
\begin{center}
\begin{tabular}{|c|c|c|c|c|}
 \hline

&&&&\\ 
Component & $V18$ & $CD-Bonn$ & $Nijmegen(93)$ & 
$Paris$  \\
&&&&\\
 \hline     
&&&&\\
 $g_\sigma$ &  9.0  
& 11.2  & 9.8 
& 7.6\\ 
&&&&\\
\hline
&&&&\\
$g_{a_0}$& 9.0 
& 9.0
& 9.0
& 10.4
\\
&&&&\\
\hline
&&&&\\
$g_\omega$& 12.2 
& 13.5
& 11.7
& 12.7
\\
&&&&\\
 \hline
&&&&\\
$\kappa_\rho$& 7.0
& 7.0
& 6.3
& 10.1
\\
&&&&\\ \hline
&&&&\\
$g_\eta$& 8.7 
& 0.0
& 1.8
& 11.7
\\
&&&&\\
 \hline
&&&&\\
$g_\pi$& 13.4 
& 13.0
& 12.7
& 13.2 
\\
&&&&\\
 \hline
\end{tabular}
\caption{Effective meson-nucleon
coupling values that correspond to
phenomenological interaction
models.}

\end{center}
\end{table}

\subsection{The vector potential}
\label{3secb}

The expression for the isospin independent 
and isospin dependent vector
potentials in configuration space are obtained from
(\ref{e11}) as
\begin{eqnarray}
&& v_V^+(r) ={1\over 4}\, \tilde v_C^+(r)
-{m_N^2\over 2}\int_r^\infty dr'\,r'\, \tilde v_{LS}^+ (r'),
\nonumber\\
&& v_V^-(r) = -{m_N^2\over 2}\int_r^\infty dr' r'\,
\tilde v_{LS}^-(r')
+{1\over 4}\{\tilde v_T^-(r)-3\int_r^\infty dr'\,
{\tilde v_T^-(r')\over r'}\} 
-{1\over 4}\,\tilde v_{SS}^-(r)
\nonumber \\
&&+{m_N^2 \over 4}\{r^2\,\tilde v_{LS2}^- (r) -3
\int_r^\infty dr' r'\, \tilde v_{LS2}^-(r')\} ,
\label{e17}
\end{eqnarray}
These potential components as obtained from the phenomenological
$V18$ \cite{V18},
the $CD-Bonn$ \cite{Mach}, the $Nijmegen(93)$ 
\cite{Swart} and the $Paris$ \cite{Paris} potentials
are shown in Figs. 3 and 4. Note that in the large
$N_C$ limit $v_V^- (r) = - v_S^- (r)$. In a boson
exchange interpretation this implies equality
in strength between the $a_0(980)$ and $\rho(770)$ exchange
interactions.

The isospin independent vector interaction admits an
obvious interpretation in terms of a strongly
repulsive $\omega$ meson exchange interaction.
The shape of this interaction component is very
similar for all the interactions considered, and
the volume integrals are also remarkably
similar, ranging from 8.7 ($Nijmegen(93)$) to
11.6 fm$^2$ ($CD-Bonn$). If the ``effective''
$\omega-$nucleon coupling constant is extracted
from the volume integrals, by setting them
equal to $g_\omega^2/m_\omega^2$ the values
range between 11.7 and 13.5.  The similarity
between these coupling constant values is
particularly notable in the case of the 
$V18$ and the parametrized $Paris$ interaction models,
as neither one of these contains any explicit
vector meson exchange term. 

The isospin dependent vector interaction, by 
eqn. (\ref{e13}), admits an interpretation in
terms of $\rho-$ meson exchange. This interaction
component is roughly 3 times weaker than the
corresponding isospin independent interaction.
The volume integrals of the isospin dependent
vector interaction range from 2.9 to 4.4 fm$^2$
for the 4 phenomenological interactions
considered here. In a simple $\rho-$meson
exchange model these volume integrals have
the form $(1+\kappa)g_\rho^2/m_\rho^2$,
where $g_\rho$ and $\kappa$ are the vector
and tensor $\rho-$meson-nucleon coupling
constants. The canonical value for the 
$\rho-$nucleon vector coupling constant
is $g_\rho^2/4\pi \sim 0.5$. If this
value for $g_\rho$ is used in the expression 
for the volume integral, an ``effective'' value
for $\kappa_\rho$ can be determined. The
volume integrals in Table 2 for the 4 interaction
models considered then give values for
$\kappa_\rho$ in the range 6.3 -- 10.1.
These values support the early finding
that $\kappa_\rho \sim 6.6$ \cite{Hohler}. 
Both the $Nijmegen93$ and the
$CD-Bonn$ interaction use the somewhat
larger value $g_\rho^2/4\pi = 0.84$ for the
$\rho$-nucleon coupling constant. This
larger value would reduce the values for
$\kappa_\rho$ in Table 3 by about 40 \%.

From eqns. (\ref{e15}) and (\ref{e17}) it follows
that in the limit of large $N_C$ the isospin
dependent scalar and vector exchange potentials
should be equal in strength and of opposite
sign. If imposed on a boson exchange model this
condition implies that
\begin{equation}
g_{a_0}^2 = (1+\kappa_\rho)g_\rho^2.
\label{e18}
\end{equation}
Comparison of the form and strength of these
potential components for the
$CD-Bonn$ and $Nijmegen(93)$ potential models
in Figs. 2 and 4 shows that these potential
models satisfy this constraint very well.

\subsection{The tensor potential}
\label{3secc}

The expression for the isospin independent 
and isospin dependent tensor
potentials in configuration space are obtained from
(\ref{e11}) as

\begin{eqnarray}
&&v_T^+(r) = -{1\over 32 m_N^2}\{\tilde v_C^{+''}(r)
+{2\over r}\,\tilde v_C^{+'}(r)\}
 + {1\over 16 }\{3\tilde v_{LS}^ +(r)+r\,\tilde v_{LS}^{+'}(r)\}
\nonumber\\
&&-{1\over 2}\{\tilde v_T^+(r)
-3\int_r^\infty dr'\,{\tilde v_T^+(r')\over r'}\} 
+{1\over 2}\tilde v_{SS}^+(r)
-{m_N^2 \over 4}\{r^2\, \tilde v_{LS2}^+ (r) -3
\int_r^\infty dr' r' \,\tilde v_{LS2}^+(r')\} ,
\nonumber\\
&&v_T^-(r) =
-{1\over 2}\{\tilde v_T^-(r)
-3\int_r^\infty dr'\,{\tilde v_T^-(r')\over r'}\} 
+{1\over 2}\tilde v_{SS}^-(r)
\nonumber\\
&&-{m_N^2 \over 4}\{ r^2\,\tilde v_{LS2}^- (r) -3
\int_r^\infty dr' r' \tilde v_{LS2}^-(r')\} ,
\label{e19}
\end{eqnarray}
These potential components as obtained from the phenomenological
$V18$ \cite{V18},
the $CD-Bonn$ \cite{Mach}, the $Nijmegen(93)$ 
\cite{Swart} and the $Paris$ \cite{Paris} potentials
are shown in Figs. 5 and 6. 

In a meson exchange model the tensor potential arises
from vector meson exchange when the vector mesons
couple to the nucleon with a Pauli (tensor)
coupling. As in the $CD-Bonn$ interaction model
the isospin independent $\omega$ meson exchange
interaction has no Pauli coupling term
$v_T^+$ vanishes in this interaction model.

By the $N_C$ counting rules in Table 1 the isospin
independent tensor interaction should be smaller
by $1/N_C^2$ than the isospin dependent
tensor interaction. Comparison of the corresponding
set of volume integrals in Table 2 shows that the
phenomenological interaction models considered
here satisfy this rule well.

The order $N_C$ of the isospin dependent tensor interaction
alone does not explain why this interaction is one
order of magnitude smaller than the corresponding vector
interactions for all the phenomenological
potential models. For boson exchange models the
reason is readily seen in eqn. (\ref{e13}), which
shows that the tensor coupled vector meson 
exchange interaction is suppressed by an overall
factor $1/m_N^2$, which is only partially
counteracted by the large tensor coupling
$\kappa_\rho$. For the phenomenological
interaction models the reason for the weakness
of the isospin dependent tensor interaction
is to be found in the fact that pion exchange
gives rise to the bulk of the isospin
dependent spin-spin and vector interactions,
and this pion (or pseudoscalar) exchange
contribution is exactly cancelled in the 
combination of $v_{SS}^-$ and $v_T^-$ in
eqn. (\ref{e19}). This latter argument may
also be extended to the isospin independent
tensor interaction, where the pseudoscalar -
in this case the $\eta$ meson - exchange
contribution cancels in the combination
of $v_{SS}^+$ and $v_T^+$ in (\ref{e19}).

\subsection{The axial vector  potential}
\label{3secd}

The expression for the isospin independent 
and isospin dependent axial vector
potentials in configuration space are obtained from
(\ref{e11}) as
\begin{eqnarray}
&&v_A^+(r) = -{1\over 32 m_N^2}\{\tilde v_C^{+''}(r)
+{2\over r}\,\tilde v_C^{+'}(r)\}
 + {1\over 16 }\{3\tilde v_{LS}^ +(r)+r\,\tilde v_{LS}^{+'}(r)\}
\nonumber\\
&&-{1\over 2}\{\tilde v_T^+(r)
-3\int_r^\infty dr'\,{\tilde v_T^+(r')\over r'}\} 
+{1\over 2}\tilde v_{SS}^+(r)
+{m_N^2 \over 4}\{r^2 \,\tilde v_{LS2}^+ (r) -3
\int_r^\infty dr' r' \,\tilde v_{LS2}^+(r')\} ,
\nonumber\\
&&v_A^-(r) =
-{1\over 2}\{\tilde v_T^-(r)
-3\int_r^\infty dr'\,{\tilde v_T^-(r')\over r'}\} 
+{1\over 2}\tilde v_{SS}^-(r)
\nonumber\\
&&+{m_N^2 \over 4}\{ r^2\, \tilde v_{LS2}^- (r) -3
\int_r^\infty dr' r'\, \tilde v_{LS2}^-(r')\} ,
\label{e20}
\end{eqnarray}
These potential components as obtained from the phenomenological
$V18$ \cite{V18},
and the $Paris$ \cite{Paris} potentials
are shown in Figs. 7 and 8.
In a single meson exchange model only axial vector
meson exchange gives rise to an axial vector
interaction. Since the $CD-Bonn$ and the
$Nijmegen93$ boson exchange interaction models do not contain
any $a_1$ meson exchange terms these interaction models
do not have any axial vector exchange components.
The phenomenological $V18$ and $Paris$ potential
models give rise to small $A$ interaction components.
These satisfy the $N_C$ counting rules in Table 1,
by which the isospin dependent axial vector
interaction should be larger by $N_C^2$ than the
isospin independent one as may be seen from the
volume integrals in Table 2. The substantial 
strength of the axial vector exchange 
interaction components of the $Paris$ potential
model are likely to be artifacts of the parametrization
rather than genuine consequences of the $NN$
scattering data.

\subsection{The pseudoscalar potential}
\label{3sece}

The expression for the isospin independent 
and isospin dependent pseudoscalar
potentials in configuration space are obtained from
(\ref{e11}) as
\begin{eqnarray}
&&v_P^+(r) = {1\over 4}\,\tilde v_C^+(r)
-{m_N^2\over 2} \int_r^\infty dr'\,r'\, \tilde v_{LS}^+ (r')
+12 m_N^2 \int_r^\infty dr' r'
\int_{r'}^\infty dr''\,{\tilde v_T^+(r'')\over r''}\, 
\nonumber\\
&&v_P^-(r) = 12 m_N^2 \int_r^\infty dr' r'
\int_{r'}^\infty dr''\,{\tilde v_T^-(r'')\over r''}\, .
\label{e21}
\end{eqnarray}
These potential components as obtained from the phenomenological
$V18$ \cite{V18},
the $CD-Bonn$ and the $Nijmegen(93)$ 
\cite{Swart} and the $Paris$ \cite{Paris} potentials
are shown in Figs. 9 and 10.

In a single meson exchange interpretation the longest range
mechanism, that contributes to the isospin independent
pseudoscalar interaction, is $\eta$ meson exchange.
This interaction component is not well
constrained by nucleon-nucleon scattering data. As a
consequence the phenomenological interaction models
considered here give widely different results for this
interaction component, as is evident in Fig. 9. 

The $\eta-$nucleon coupling constant is also not well
known. An estimate for this coupling constant may be 
obtained from the volume integrals of the phenomenological
interaction models if these are set to equal
$g_\eta^2/m_\eta^2$. From the volume integral values
listed in Table 2 one then obtains values for
$g_\eta$, which range from 1.8 ($Nijmegen(93)$) to
11.7 ($Paris$) (Table 3). Analyses of observables,
other than nucleon-nucleon scattering, as e.g.
$\eta-$meson photoproduction suggest that the
coupling constant value should not exceed 2.2 \cite{Benn}. 

The isospin dependent pseudoscalar exchange interaction
is strong and has long range. Its main component is the
long range pion exchange interaction, which is built into
all the phenomenological interaction models considered
here. Because of this all the interaction models converge
for nucleon separations larger than 1 fm. This 
interaction component is also the strongest component
by $N_C$ counting, as it scales as $N_C^3$ (Table 1).
The volume integrals of this interaction component
are listed in Table 2 for the 4 interaction models
considered, and range from 323 fm$^2$ to 360 fm$^2$.
The corresponding values for the pseudoscalar
pion-nucleon coupling constant $g_\pi$ range from
12.7 to 13.4, when extracted by equating the
numerically determined volume integrals with
$g_\pi^2/m_\pi^2$ (Table 3).

\section{Electromagnetic exchange current in the 
large ${\bf N_C}$ limit}
\label{4sec}

	The isoscalar and isovector components
of the single nucleon charge operator are of order
$N_C^1$ and $N_C^0$ respectively, as may be inferred
from the scaling relations (\ref{e9}). 
Because of their inverse dependence on
the nucleon mass the isoscalar and isovector
convection current components of the
single nucleon current scale as $N_C^0$ and
$N_C^{-1}$ in comparison. In the case of the spin
component of the single nucleon current
the isoscalar component scales in contrast
as $N_C^{-1}$, whereas the $N_C$ dependence of
the isovector part is $N_C^0$.

Associated with the isospin dependent interactions expressed in
terms of Fermi invariants are corresponding two-nucleon
electromagnetic interaction currents with the isospin dependence 
$(\vec\tau^1\times\vec\tau^2)_3$ \cite{Tsushima}. In a meson
exchange interpretation, these interaction currents
correspond to the currents carried by exchanged charged
mesons (Fig. 11b). These current operators are listed in Table 4,
along with the corresponding $N_C$ scaling power.
The $N_C$ scaling factors of these
current operators may be derived by reduction
to the spin representation, and application
of the relations (\ref{e9}).
In the expressions in Table 4 the potential
functions $v_j^\pm (k) $ represent the Fermi invariant
potential components in momentum space.
The $N_C$ dependence of these interaction
currents should be compared to the $N_C$
scaling of the corresponding components of the 
single nucleon charge and current operators.

\begin{table}
\begin{center}
\begin{tabular}{|c|c|c|}
 \hline

&&\\ 
Fermi invariant & $j_\mu$ & $N_C$   \\
&&\\
 \hline     
&&\\
 $S$ & $ i(\vec\tau^1\times\vec\tau^2)_3
{v_S^-(k_2)-v_S^-(k_1)\over k_2^2-k_1^2} 
 (k_{2\mu}-k_{1\mu})S$
& $N_C^{-1}$
\\ 
&&\\
\hline
&&\\
$V$&  $\{ i(\vec\tau^1\times\vec\tau^2)_3
{v_V^-(k_2)-v_V^-(k_1)\over k_2^2-k_1^2}$ 
& \\
&&\\
&$ [(k_{2\mu}-k_{1\mu})V+\gamma_\mu^1(\gamma^2\cdot k_1)
-\gamma_\mu^2(\gamma^1\cdot k_2)]\}$ &$N_C^{-1}$\\
&&\\
\hline
&&\\
$T$&  $\{ i(\vec\tau^1\times\vec\tau^2)_3
{v_T^-(k_2)-v_T^-(k_1)\over k_2^2-k_1^2}$ 
& \\
&&\\
&$ [(k_{2\mu}-k_{1\mu})T+k_{2\nu}\sigma_{\mu\alpha}^2
\sigma_{\alpha\nu}^1
-k_{1\nu}\sigma_{\mu\alpha}^1
\sigma_{\alpha\nu}^2]\}$ &$N_C^1$\\
&&\\

\hline
&&\\
$A$& $ i(\vec\tau^1\times\vec\tau^2)_3
{v_A^-(k_2)-v_A^-(k_1)\over k_2^2-k_1^2} 
 (k_{2\mu}-k_{1\mu})A$
& $N_C^1$

\\
&&\\ \hline
&&\\
$P$&  $ i(\vec\tau^1\times\vec\tau^2)_3
\quad{v_P^-(k_2)-v_P^-(k_1)\over k_2^2-k_1^2} 
 (k_{2\mu}-k_{1\mu})P$
& $N_C^{-1}$\\
&&\\
 \hline
\end{tabular}
\caption{Interaction currents and
$N_C$ scaling factors associated with the
interaction components. The fractions of the
imparted momentum $q$ to the two nucleons are
denoted $k_1$ and $k_2$. Note that in the
case of the $A$ invariant a conserved
current obtains only in combination
with a P invariant term \cite{Tsushima}.  }

\end{center}
\end{table}

The charge components of the interaction
currents in Table 4 are suppressed by their
dependence on the energy exchanged between
the nucleons. This is proportional to the
nucleon momenta and inversely proportional
to the nucleon mass ($\sim 1/N_C$). In a
frame with no energy transfer they vanish.
In the reduction
to the spin representation exchange charge
operators also arise in the elimination of negative
energy intermediate states (``pair terms''). These 
are listed in Table 5. These operators have an overall
factor $1/m_N^3$, which arises from the
small components of the Dirac spinors and the
propagator of the intermediate negative energy
propagator (Fig. 11a). The order in $1/N_C$ of these
operators follow from the general scaling
rules (\ref{e9}) and the overall factor
$1/m_N^3 \sim 1/N_C^3$. In Table 5 and
in subsequent tables, the momenta of the
two nucleons are denoted by
$\vec P_j=1/2(\vec p_j\,'+\vec p_j)$
with $j=1,2$ respectively.

	The largest $N_C$ scaling factors of the
exchange charge operators that are associated
with the 5 Fermi invariant components of the
interaction are also listed in Table 5, separately
for the isospin independent and isospin
dependent interaction components. The order of the 
exchange charge operators are smaller by $1/N_C^3$
in comparison with the corresponding
single nucleon charge
operators. The terms with the highest $N_C$
dependence are those, which are independent
of spin and isospin and those, which 
contain the spin-isospin bilinears
$\sigma_i\tau_j$ of both nucleons. The remaining
terms are smaller by at least one power of
$1/N_C$.

\begin{table}
\begin{center}
\begin{tabular}{|c|c|c|c|}
 \hline

&&&\\ 
Fermi invariant & $\rho$ & $v_j^+$ & $v_j^-$  \\
&&&\\
 \hline     
&&&\\
 $S$ & $ {1\over 4 m_N^3}
[\vec q\,^2+2i \vec\sigma^1\cdot \vec P_1\times\vec q]
$&&\\
&&&\\
&$[v_S^+ (k_2)\hat e_1+v_S^- (k_2)\tilde e_{2}]
+(1\leftrightarrow 2)$
&$N_C^{-2}$ 
&$N_C^{-3}$ 
\\ 
&&&\\
\hline
&&&\\
$V$&   $ {1\over 4 m_N^3}
[\vec q\cdot\vec k_2+(\vec\sigma^2\times\vec k_2)
\cdot(\vec\sigma^1\times\vec q)-2i\vec\sigma^1\cdot \vec q
\times \vec P_2]$ 
&& \\
&&&\\
&$[v_V^+ (k_2)\hat e_1+v_V^- (k_2)\tilde e_2]
+(1\leftrightarrow 2)$& $N_C^{-2}$&$N_C^{-2}$\\
&&&\\
\hline
&&&\\
$T$&  $ {1\over 4 m_N^3}
[\vec\sigma^1\cdot\vec\sigma^2\, \vec q\cdot\vec k_2
-\vec\sigma^2\cdot\vec q\, \vec\sigma^1\cdot
(\vec k_2-\vec q)$ & 
& \\
&&&\\
&$+\vec q\cdot \vec k_2 -2 i\vec \sigma^2\cdot \vec q\times
\vec P_2]$&&\\
&&&\\
&$ [v_T^+ (k_2)\hat e_1+v_T^- (k_2)\tilde e_2]
+(1\leftrightarrow 2)$ 
&$N_C^{-2}$&$N_C^{-2}$\\
&&&\\
\hline

&&&\\
$A$& $  {1\over 4 m_N^3}
[\vec q\, ^2\vec \sigma^1\cdot\vec \sigma^2+2i\vec\sigma^2\cdot
\vec P_1\times \vec q$
&& \\
&&&\\
&$-\vec \sigma^1\cdot \vec q\, \vec\sigma^2\cdot\vec q
+\vec\sigma^1\cdot \vec q\,\vec\sigma^2\cdot\vec k_2]$
&&\\
&&&\\
&$ [v_A^+ (k_2)\hat e_1+v_A^- (k_2)\tilde e_2]
+(1\leftrightarrow 2)$ 
&$N_C^{-2}$&$N_C^{-2}$\\

&&&\\ \hline

&&&\\
$P$&  $ {1\over 4 m_N^3}
\vec\sigma^2\cdot\vec k_2\, \vec\sigma^1\cdot\vec q
$ &&\\
&&&\\
&$[v_P^+ (k_2)\hat e_1+v_P^- (k_2)\tilde e_2]
+(1\leftrightarrow 2)$
& $N_C^{-3}$&$N_C^{-2} $,\\
&&&\\                     
 \hline
\end{tabular}
\caption{Exchange charge operators and
the highest $N_C$ scaling factors associated with the
interaction components. The fractions of the
imparted momentum $q$ to the two nucleons are
denoted $k_1$ and $k_2$ respectively. The
isospin factors are defined as
$\hat e_1=(1+\tau_3^1)/2$ and
$\tilde e_2 =(\vec\tau^1\cdot\vec\tau^2+\tau_3^2)/2$. }

\end{center}
\end{table}

The interaction currents that are listed
in Table 4 arise from the coupling of the
electromagnetic field to the charge exchanged
between the nucleons (Fig. 11b). Their 
explicit expressions
in the spin representation have been
given in ref. \cite{Tsushima}. The components
of these currents, which are of highest order
in $N_C$ are listed in Table 6.

\begin{table}
\begin{center}
\begin{tabular}{|c|c|c|}
 \hline

&&\\ 
Fermi invariant & $\vec j $ & ${\cal O}(N_C)$   \\
&&\\
 \hline     
&&\\
 $S$ & $ i(\vec\tau^1\times\vec\tau^2)_3
{v_S^-(k_2)-v_S^-(k_1)\over k_2^2-k_1^2} 
 (\vec k_2-\vec k_1)$
& $N_C^{-1}$
\\ 
&&\\
\hline
&&\\
$V$&  $\{ i(\vec\tau^1\times\vec\tau^2)_3
{v_V^-(k_2)-v_V^-(k_1)\over k_2^2-k_1^2}$ 
& \\
&&\\
&$\{ (\vec k_2-\vec k_1)\{1+{1\over 4 m_N^2}
[\vec\sigma^1\cdot\vec \sigma^2\,\vec k_1\cdot\vec k_2
-\vec\sigma^1\cdot \vec k_2\,\vec\sigma^2\cdot
\vec k_1]$ &\\ 
&&\\
&$-{1\over 4 m_N^2}[\vec\sigma^1\times\vec k_1\,
\vec\sigma^2\cdot\vec k_1\times\vec k_2
+\vec\sigma^2\times\vec k_2 \,
\vec\sigma^1\cdot\vec k_1\times\vec k_2]\}
$ & $N_C^{-1}$\\
&&\\
\hline
&&\\
$T$&  $\{ i(\vec\tau^1\times\vec\tau^2)_3
{v_T^-(k_2)-v_T^-(k_1)\over k_2^2-k_1^2} 

 [\vec\sigma^1\,\vec\sigma^2\cdot \vec k_2 
-\vec\sigma^2\,\vec\sigma^1\cdot\vec k_1] $ &$N_C^1$\\
&&\\

\hline
&&\\
$A$& $ i(\vec\tau^1\times\vec\tau^2)_3
{v_A^-(k_2)-v_A^-(k_1)\over k_2^2-k_1^2} 
 (\vec k_2-\vec k_1)\,\vec\sigma^1\cdot\vec \sigma^2$
& $N_C^1$

\\
&&\\ \hline
&&\\
$P$&  $ {i\over 4 m_N^2}(\vec\tau^1\times\vec\tau^2)_3
\quad{v_P^-(k_2)-v_P^-(k_1)\over k_2^2-k_1^2} 
 (\vec k_2-\vec k_1)\,\vec\sigma^1\cdot \vec k_1\,
\vec\sigma^2\cdot \vec k_2$
& $N_C^{-1}$\\
&&\\
 \hline
\end{tabular}
\caption{Interaction 
currents that are associated with the Fermi
invariant interaction components that arise
from e.m. coupling to the exchanged charge,
which have the highest
$N_C$ scaling factors associated with the
interaction components. The fractions of the
imparted momentum $q$ to the two nucleons are
denoted $k_1$ and $k_2$ respectively.   }

\end{center}
\end{table}

The $N_C$ dependence of these internal coupling
interaction currents that are associated with
the different Fermi invariant components of
the phenomenological interaction model
is smaller by one power of $1/N_C$ than
that of the isovector
part of the single nucleon current, with 
exception for the currents that are
associated with the $T$ and $A$ invariants,
which anomalously scale as $N_C^1$.
As these interaction components are very
weak in all the phenomenological models
considered, and are suppressed by 
$1/m_N^2\sim 1/N_C^2$ in
meson exchange models, it is natural to
conclude that their $N_C$ dependence in
practice also is $N_C^{-1}$ rather than
$N_C^1$.

In the reduction to the spin representation
the negative energy pole terms give rise
to seagull type exchange currents, in which the
electromagnetic field couples to either one of
the interaction nucleons with a point interaction
to the interaction (Fig. 11a). The exchange currents of
this type, which have the highest $N_C$ scaling
factors (in this case $1/N_C$) are listed in Table 7.
The $1/N_C$ dependence of these current
operators are similar to the internal
charge coupling currents in Table 6.

\begin{table}
\begin{center}
\begin{tabular}{|c|c|c|}
 \hline

&&\\ 
Fermi invariant & $\vec j $ & ${\cal O}(N_C)$   \\
&&\\
 \hline     
&&\\
 $S$ & $- {1\over 4 m_N^2}(1+\vec\tau^1_3)
v_S^+ (k_2)(i\vec \sigma^1\times\vec q+2 \vec P_1)
+(1\leftrightarrow 2)$
& $N_C^{-1}$
\\ 
&&\\
\hline
&&\\
$V$&  $- {1\over 4 m_N^2}\{(1+\vec\tau^1_3)
v_V^+ (k_2)(i(\vec \sigma^1+\vec\sigma^2)\times \vec k_2
+2\vec P_2)$
& \\
&&\\
& $+i(\vec\tau^1\times\vec\tau^2)_3 v_V^-(k_2)
[\vec k_2\,\vec\sigma^1\cdot\vec\sigma^2
-\vec\sigma^2\,\vec\sigma^1\cdot \vec k_2]\} 
+(1\leftrightarrow 2)$ 
& $N_C^{-1}$\\
&&\\
\hline
&&\\
$T$&  $- {1\over 4 m_N^2}\{i(1+\vec\tau^1_3)
v_T^+ (k_2)(\vec \sigma^1+\vec\sigma^2)\times \vec k_2
$&\\
&&\\
&$+(\vec\tau^1\cdot\vec\tau^2+\vec \tau^2_3)
v_T^- (k_2)[2\vec\sigma^1\,\vec\sigma^2\cdot\vec P_1
-2\vec\sigma^2\,\vec\sigma^1\cdot\vec P_2
+2\vec P_2\sigma^1\cdot\vec\sigma^2]
$&\\
&&\\
&$+i(\vec\tau^1\times\vec\tau^2)_3 v_T^- (k_2)
[\vec\sigma^1\,\vec\sigma^2\cdot \vec q
+\vec k_2\,\vec\sigma^1\cdot\vec\sigma^2
-\vec\sigma^2\,\vec\sigma^1\cdot\vec k_2]\}
+(1\leftrightarrow 2)
 $ &$N_C^{-1}$\\
&&\\

\hline
&&\\
$A$&  $- {1\over 4 m_N^2}\{(\vec\tau^1\cdot\vec\tau^2
+\vec \tau^2_3)v_A^- (k_2)
[2\vec\sigma^1\,\vec\sigma^2\cdot \vec P_2
-2\vec\sigma^2\,\vec\sigma^1\cdot \vec P_1
+2\vec P_1\,\vec\sigma^1\cdot\vec\sigma^2]
$&\\
&&\\
&$+i(\vec\tau^1\times\vec\tau^2)_3 v_A^- (k_2)
[\vec q \,\vec\sigma^1\cdot\vec\sigma^2
-\vec \sigma^2\,\vec\sigma^1\cdot\vec q
+\vec \sigma^1\,\vec \sigma^2\cdot \vec k_2]\}
+(1\leftrightarrow 2)
$&$N_C^{-1}$\\
&&\\ \hline

&&\\
$P$&  $- {i\over 4 m_N^2}
(\vec\tau^1\times\vec\tau^2)_3 v_P^- (k_2)
\vec\sigma^1\,\vec\sigma^2\cdot\vec k_2
+(1\leftrightarrow 2)
$&$N_C^{-1}$\\
&&\\

 \hline
\end{tabular}
\caption{Interaction 
currents that arise from coupling to intermediate
negative energy pole terms that have the largest
$N_C$ scaling factors. The fractions of the
imparted momentum $q$ to the two nucleons are
denoted $k_1$ and $k_2$ respectively.   }

\end{center}
\end{table}

\section{Axial exchange current in the large 
${\bf N_C}$ limit}
\label{5sec}

The axial current operator of a single nucleon,
\begin{equation}
\vec A_\pm=-g_A\vec\sigma\tau_\pm,
\label{e22}
\end{equation}
is of order $N_C^1$ (cf. (\ref{e9})).
The axial vector interaction currents that are
associated with the Fermi invariant components
of the nucleon-nucleon interaction have been
listed in ref. \cite{Tsu2}. These currents only arise
in the reduction to the spin representation,
which brings in an overall factor $1/M_N^3$. The
$1/N_C$ scaling factors of these axial exchange
current operators may be determined from their
operator structure by means of the relations
(\ref{e9}). It follows from their vector nature
that they are $1/N_C^3$ 
smaller than those of the corresponding
interaction components. 

	The subset of axial exchange currents that are
associated with the Fermi invariant interaction
components, and which are
of order $1/N_C^2$ are listed in Table 8. In the case of the
pseudoscalar invariant the reduction to
the spin representation brings an overall
factor $1/M_N^5$. The axial exchange current 
that is associated
with the pseudoscalar invariant is therefore 
of the order $N_C^{-5}$.

\begin{table}
\begin{center}
\begin{tabular}{|c|c|c|}
 \hline

&&\\
Fermi invariant & $\vec A$ & ${\cal O}(N_C)$  \\
&&\\
 \hline     
&&\\
$S$&$-{g_A\over 4 m_N^3}v_S^+(k_2)\,\tau_\pm^1
(4\vec\sigma^1 \vec P_1^2
-4\vec P_1\vec\sigma^1\cdot\vec P_1)
+(1\leftrightarrow 2)
$&$1/N_C^2$\\
&&\\
\hline
&&\\
$V$&$-{g_A\over 4 m_N^3}\{v_V^+(k_2)\,\tau_\pm^1
\,[\,{3\over 4}\vec\sigma^1 \vec k_2^2
-\vec\sigma^1\vec P_1\,^2 -\vec k_2\vec\sigma^1\cdot \vec k_2
$&\\
&&\\
&$+4\vec\sigma^1 \vec P_1\cdot\vec P_2
-4\vec P_2\vec\sigma^1\cdot\vec P_1]
+v_V^-(k_2)\tau_\pm^2[\vec\sigma^2\vec k_2^2
-\vec k_2\vec\sigma^2\cdot\vec k_2]
$&\\
&&\\
&$+v_V^-(k_2)(\vec\tau^1\times\vec\tau^2)_\pm 
\vec\sigma^2\times\vec k_2\,\vec\sigma^1\cdot\vec k_2\}
+(1\leftrightarrow 2)$
&$1/N_C^2$\\
&&\\
\hline
&&\\
$T$&$-{g_A\over 4 m_N^3}\{v_T^+(k_2)\,\tau_\pm^1
\,(\vec\sigma^1 \vec k_2^2-\vec k_2\vec\sigma^1
\cdot\vec k_2)
$&\\
&&\\
&$+v_T^-(k_2)\tau_\pm^2\,[ 4 \vec\sigma^2\vec P_1\cdot\vec P_2
-4\vec P_2\vec\sigma^2\cdot\vec P_1
-\vec\sigma^2\vec P_1^2+{3\over 4}\vec\sigma^2 \vec k_2^2
-\vec k_2\vec\sigma^2\cdot\vec k_2]
$&\\
&&\\
&$-v_T^-(k_2)(\vec\tau^1\times\vec\tau^2)_\pm 
[4\vec\sigma^1\vec P_1\cdot\vec\sigma^2\times\vec P_2
-4\vec\sigma^2\times\vec P_2 \vec\sigma^1\cdot\vec P_1
$&\\
&&\\
&$+3\vec\sigma^1\times\vec P_1\vec\sigma^2\cdot\vec P_1
-{1\over 4}\vec\sigma^1\times\vec k_2\vec \sigma^2
\cdot \vec k_2+\vec\sigma^2\times\vec P_1\vec\sigma^2
\cdot\vec P_1
$&\\
&&\\
&$-{3\over 4}\vec\sigma^2\times\vec k_2
\vec\sigma_1\cdot\vec k_2
-\vec P_1\vec\sigma^1\times\vec\sigma^2\cdot\vec P_1
-{1\over 4}\vec k_2\vec\sigma^1\times\vec\sigma^2
\cdot \vec k_2\}+(1\leftrightarrow 2)
$&$1/N_C^2$\\
&&\\
\hline
&&\\
$A$&$-{g_A\over 4 m_N^3}\{v_A^-(k_2)\,\tau_\pm^2
\,(3\vec\sigma^2\vec P_1^2
-{1\over 4} \vec\sigma^2\vec k_2^2
-4\vec P_1\vec\sigma^2\cdot\vec P_1
$&\\
&&\\
&$-v_A^-(k_2)(\vec\tau^1\times\vec\tau^2)_\pm 
[4\vec\sigma^1\times\vec P_1\vec\sigma^2\cdot\vec P_2
-\vec\sigma^1\times\vec P_1\vec\sigma^2\cdot\vec P_1
$&\\
&&\\
&$
+{3\over 4}\vec\sigma^1\times\vec k_2 
\vec\sigma^2\cdot\vec k_2
-3\vec\sigma^2\times\vec P_1\vec\sigma^1\cdot\vec P_1
+{1\over 4}\vec\sigma^2\times\vec k_2\vec \sigma^1
\cdot \vec k_2

$&\\
&&\\
&$-\vec P_1 \vec\sigma^1\times\vec\sigma^2
\cdot\vec P_1
-{1\over 4}\vec k_ 2\vec\sigma^1\times
\vec\sigma_2\cdot\vec k_2\}+(1\leftrightarrow 2)
$&$1/N_C^2$\\
&&\\
\hline

\end{tabular}
\caption{Axial exchange current operators with the
highest $N_C$ scaling factors associated with the
interaction components. }

\end{center}
\end{table}

	The axial exchange charge operators
that are associated with Fermi invariant
decomposition of the nucleon-nucleon
interaction have been derived in ref. \cite{Mariana},
and are listed in Table 9.
For these operators the reduction to the
spin representation only brings in an
overall factor of $1/m_N^2$. The $1/N_C$
scaling factors of these operators are
accordingly smaller by a factor
$1/N_C^2$ than those of the corresponding
interaction components. 

	The axial charge operator of a single
nucleon
\begin{equation}
A_\pm^0=-g_A{\vec\sigma\cdot\vec p\over m_N}\tau_\pm,
\label{e23}
\end{equation}
is readily seen to be of order $N_C^0$ by
means of the relations (\ref{e9}). The
main axial exchange charge operators
listed in Table 6 are in comparison
of order $N_C^{-1}$.
This may be seen by referring to
the operator scaling relations (\ref{e9}).

\begin{table}
\begin{center}
\begin{tabular}{|c|c|c|c|}
 \hline

&&&\\
Fermi invariant & $A^0$ & $v_j^+$ & $v_j^-$  \\
&&&\\
 \hline     
&&&\\
 $S$ & $ {g_A\over m_N^2}
[v_S^+ (k_2)\tau_\pm^1+v_S^- (k_2)\tau_\pm^2]
\vec\sigma^1\cdot \vec P_1
+(1\leftrightarrow 2)
$& $N_C^{-1}$&$N_C^{-3}$
\\ 
&&&\\
\hline
&&&\\
$V$&   $ {g_A\over m_N^2}
\{[v_V^+ (k_2)\tau_\pm^1+v_V^- (k_2)\tau_\pm^2]
[\vec\sigma^1\cdot\vec P_2+{1\over 2}
\vec\sigma^1\times\vec\sigma^2\cdot \vec k_2]$
&&\\
&&&\\
&$+{i\over 2}v_V^- (k_2)(\vec\tau^1\times\vec\tau^2)_\pm
\vec\sigma^1\cdot \vec k_2\}
+(1\leftrightarrow 2)$& $N_C^{-1}$&$N_C^{-2}$\\
&&&\\
\hline
&&&\\
$T$&  $ {g_A\over m_N^2}
\{[v_T^+ (k_2)\tau_\pm^1+v_T^- (k_2)\tau_\pm^2]
[\vec\sigma^2\cdot\vec P_1+{i\over 2}
\vec\sigma^1\times\vec\sigma^2\cdot\vec k_2]
$ & 
& \\
&&&\\
&$
+iv_T^- (k_2)(\vec\tau^1\times\vec\tau^2)_\pm
[{1\over 2}\vec\sigma^1\cdot \vec k_2+i\vec\sigma^1
\times\vec\sigma^2\cdot\vec P_2]\}+(1\leftrightarrow 2)$ 
&$N_C^{-2}$&$N_C^{-1}$\\
&&&\\
\hline

&&&\\
$A$& $  {g_A\over m_N^2}\{
[v_A^+ (k_2)\tau_\pm^1+v_A^- (k_2)\tau_\pm^2]
\vec\sigma^2\cdot \vec P_2$
&& \\
&&&\\
&$+iv_A^- (k_2)(\vec\tau^1\times\vec\tau^2)_\pm
[{1\over 2}\vec \sigma^2\cdot \vec k_2
+i\vec\sigma^1\times\vec\sigma^2\cdot \vec P_1
] \}
+(1\leftrightarrow 2)$ 
&$N_C^{-3}$&$N_C^{-1}$\\

&&&\\ \hline

\end{tabular}
\caption{Axial exchange charge operators and
highest $N_C$ scaling factors associated with the
interaction components. The fractions of the
imparted momentum $q$ to the two nucleons are
denoted $k_1$ and $k_2$ respectively. }

\end{center}
\end{table}

There is no axial exchange charge operator
that would be directly proportional to
the pseudoscalar invariant potential
component. There is however an axial
exchange charge operator, which is of
order $N_C^0$, and which arises from the
long range pion exchange interaction,
which gives rise to the bulk of the
pseudoscalar interaction component.
This axial charge operator was derived
in ref. \cite{Kubodera}, and has the form
\begin{equation}
A_\pm^0(\pi)={g_A\over 4\pi} 
({m_\pi\over 2 f_\pi})^2(\vec\sigma^1
+\vec\sigma^2)\cdot\vec r_{12}
(\vec\tau^1\times\vec\tau^2)_\pm
(1+{1\over m_\pi r_{12}})
{e^{-m_\pi r_{12}}\over m_\pi r_{12}}.
\label{e24}
\end{equation}
Because the pion decay constant
$f_\pi\sim N_C$ in the large $N_C$ limit
it is readily seen by the relations (\ref{e9}) 
that this operator is of order $N_C^0$.

The contributions of the two-nucleon
exchange current terms to the axial
charge operator are conventionally
expressed in the from of an effective
single nucleon axial charge operator.
The exchange current contributions then
represent nuclear enhancement factors
of the single nucleon axial charge
operator (\ref{e23}). Experimentally this
enhancement factor is a factor 2 as
measured by first forbidden $\beta$ transitions
of nuclei in the lead region \cite{warb}. About
one half of this enhancement may be attributed
to the pion exchange operator (\ref{e24}),
which is natural as its order in the $1/N_C$
expansion is the same as that of the single
nucleon operator (\ref{e23}). The
bulk of the remainder arises from
the isospin independent scalar contribution
in Table 9 \cite{Mariana}. While the order of this term
is $1/N_C$, it gives rise to a direct term
matrix element, which involves an additional factor
of $N_C$ in the nucleon density, and therefore its
contribution to the ``effective'' single nucleon
matrix element is also of order $N_C^0$.

\section{Discussion}
\label{6sec}

This analysis of 4 commonly employed phenomenological
nucleon-nucleon interaction models reveals that the
structure of their components
is completely consistent with their corresponding
dependence on $N_C$ (or $1/N_C$). 
For all the interaction models the isospin
dependent pseudoscalar components, which
contain the long range pion exchange interaction,
have volume integrals that are larger than
those of any other interaction component by
more than an order of magnitude. This
interaction component also scales with the
largest power of $N_C$ ($N_C^3$) (Table 2).
As this interaction component shows very little
variation between 3 of the 4 interaction
models considered, it may be viewed as
well determined, if not completely settled
in form (Fig. 10).

The interaction components that follow the
isospin dependent pseudoscalar interaction
in strength are the isospin independent
scalar and vector interactions, which
scale as $N_C$. These interaction components
are those responsible for nuclear binding
($v_S^+$) and short range repulsion 
between nucleons ($v_V^-$). For distances
shorter than 0.4 fm the variation in form
of these interaction components between the  
4 considered phenomenological interaction
models is substantial. 

Next in strength are the isospin dependent
scalar and vector interactions, which also
scale as $N_C$. The volume integrals of these
interaction components are about half
as large as those of the corresponding
isospin independent interactions. A new
finding is that these two interaction
components have to have equal magnitude
and opposite sign in the large $N_C$ limit.

The tensor and axial vector interaction
components are very weak for all the
considered phenomenological interaction
models. For the isospin independent 
tensor and axial vector components this
is completely consistent with their
$N_C$ dependence, which is $1/N_C$. 
The smallness of the isospin dependent
tensor and axial vector exchange
interaction components cannot be explained by
their $N_C$ dependence alone, as they
scale as $N_C$. In meson exchange
models the smallness of $v_T^-$ is however
natural, as it arises from tensor
coupled vector mesons, and contains
an overall factor $1/m_N^2$ (\ref{e13}).

The least well understood interaction
component is the isospin independent
pseudoscalar interaction (Fig. 9). This
component, which scales as $N_C^1$,    
is vanishingly small in two of the
considered phenomenological interactions
and even stronger than $v_V^+$ in the
other two. The longest range part of
this interaction component arises from
$\eta$ meson exchange. The large
variation between the phenomenological
interaction models for this interaction
component reflects the continuing 
uncertainty concerning the strength of
the $\eta-$nucleon coupling. The $V18$
and $Paris$ interaction models have
strong ``effective'' $\eta$ nucleon
couplings, which correspond to the
old pseudoscalar $SU(3)$ coupling
model by which 
$g_{\eta NN}=g_{\pi NN}/\sqrt{3}$.

The present results for the $N_C$
dependence of the exchange currents
follow those of the corresponding
interactions. In the case of the
electromagnetic interaction current the
result that they are smaller
by $1/N_C^2$ than the corresponding
interactions fits with the phenomenological
finding that electromagnetic exchange
current contributions to nuclear
observables typically represent
effects of the order 10\%, the 
exceptions being situations where
single nucleon current matrix
elements are suppressed \cite{dor}.

The interaction current contributions to
the axial charge operator are known
to be significant in comparison with
the axial charge operators of the
single nucleons \cite{warb}. This
feature also emerges from the $1/N_C$
expansion, as the axial exchange
charge operators scale in $1/N_C$ as
the single nucleon charge operator,
or with one more power of $1/N_C$.

The overall conclusion is thus that the
ordering of nuclear interactions and
interaction current operators 
by their dependence on $1/N_C$ corresponds
well with their significance ordering
based on empirical evidence and
phenomenlogical analysis.

\vspace{1cm}
\centerline{\bf Acknowledgments}
\vspace{1cm}
The hospitality of the W. K. Kellogg
Radiation Laboratory of the California Institute
of Technology is gratefully acknowledged. 
I am indebted to C. Helminen for a careful
reading of the manuscript.
Research supported in part by the Academy of Finland
by grant 54038. 

%\newpage

\vspace{1cm}

\centerline{Figure captions}

\vspace{0.5cm}

Fig. 1 Isospin independent scalar potential coefficients
for the $V18$ \cite{V18},
the $CD-Bonn$ \cite{Mach}, the $Nijmegen(93)$ 
\cite{Swart} and the $Paris$ \cite{Paris} interaction
models.

\vspace{0.5cm}

Fig. 2 Isospin dependent scalar potential coefficients
for the $V18$ \cite{V18},
the $CD-Bonn$ \cite{Mach}, the $Nijmegen(93)$ 
\cite{Swart} and the $Paris$ \cite{Paris} interaction
models.

\vspace{0.5cm}

Fig. 3 Isospin independent vector potential coefficients
for the $V18$ \cite{V18},
the $CD-Bonn$ \cite{Mach}, the $Nijmegen(93)$ 
\cite{Swart} and the $Paris$ \cite{Paris} interaction
models.

\vspace{0.5cm}

Fig. 4 Isospin dependent vector potential coefficients
for the $V18$ \cite{V18},
the $CD-Bonn$ \cite{Mach}, the $Nijmegen(93)$ 
\cite{Swart} and the $Paris$ \cite{Paris} interaction
models.

\vspace{0.5cm}

Fig. 5 Isospin independent tensor potential coefficients
for the $V18$ \cite{V18}, the $Nijmegen(93)$ 
\cite{Swart} and the $Paris$ \cite{Paris} interaction
models.

\vspace{0.5cm}

Fig. 6 Isospin dependent tensor potential coefficients
for the $V18$ \cite{V18},
the $CD-Bonn$ \cite{Mach}, the $Nijmegen(93)$ 
\cite{Swart} and the $Paris$ \cite{Paris} interaction
models.

\vspace{0.5cm}

Fig. 7 Isospin independent axial vector potential coefficients
for the $V18$ \cite{V18} 
and the $Paris$ \cite{Paris} interaction
models.

\vspace{0.5cm}

Fig. 8 Isospin dependent axial vector potential coefficients
for the $V18$ \cite{V18}, the $Nijmegen(93)$ 
and the $Paris$ \cite{Paris} interaction
models.

\vspace{0.5cm}

Fig. 9 Isospin dependent pseudoscalar potential coefficients
for the $V18$ \cite{V18}, the $Nijmegen(93)$ 
\cite{Swart} and the $Paris$ \cite{Paris} interaction
models.

\vspace{0.5cm}

Fig. 10 Isospin dependent pseudoscalar potential coefficients
for the $V18$ \cite{V18},
the $CD-Bonn$ \cite{Mach}, the $Nijmegen(93)$ 
\cite{Swart} and the $Paris$ \cite{Paris} interaction
models.

\vspace{0.5cm}

Fig. 11 (a) Point coupling of an extrernal field
to one nucleon and a
nucleon-nucleon interaction, (b) Coupling to the
charge exchanged by a nucleon-nucleon interaction.

\end{document}